\documentclass[referee]{cjaa}           % referee version: for submission

\usepackage{graphicx}                   %for PS/EPS graphics inclusion, new
\input{epsf.sty}                        %for PS/EPS graphics inclusion, old
\input{psfig.sty}                       %for PS/EPS graphics inclusion, old

\begin{document}

   \title{Two novel approaches for photometric redshift estimation based on SDSS and 2MASS databases\,$^*$
%\footnotetext{$*$ Supported by the National Natural Science Foundation of China.}
}

   \volnopage{Vol.0 (200x) No.0, 000--000}      %%preserved for Editor. DOn't remove!
   \setcounter{page}{1}          %%starting page, preserved for Editor. DOn't remove!

   \author{Dan Wang \inst{1,2}\mailto{1} \and Yan-Xia Zhang \inst{1} \and Chao Liu \inst{1,2}
   \and Yong-Heng Zhao
      \inst{1}
      }
  % \offprints{Y. Zhang}
   \institute{National Astronomical Observatories, Chinese Academy of Sciences,
             Beijing 100012\\
             \email{dwang@lamost.org}
              \and Graduate University of Chinese Academy of Sciences, Beijing 100080,
China\\}

   \date{Received~~2007.3; accepted~~}

\abstract{We investigate two training-set methods: support vector
machines (SVMs) and Kernel Regression (KR) for photometric redshift
estimation with the data from the Sloan Digital Sky Survey Data
Release 5 and Two Micron All Sky Survey databases. We probe the
performances of SVMs and KR for different input patterns. Our
experiments show that the more parameters considered, the accuracy
doesn't always increase, and only when appropriate parameters
chosen, the accuracy can improve. Moreover for the two approaches,
the best input pattern is different. With various parameters as
input, the optimal bandwidth is dissimilar for KR. The rms errors of
photometric redshifts based on SVM and KR methods, are less than
0.03 and 0.02, respectively. Finally the strengths and weaknesses of
these techniques are summarized. Compared to other methods of
estimating photometric redshifts, they show their superiorities,
especially KR, in terms of accuracy. }

   \authorrunning{D. Wang, Y. X. Zhang, C. Liu, \& Y. H. Zhao }            %author_head in even pages
   \titlerunning{Two novel approaches for photometric redshift estimation}  % title_head in odd pages

   \maketitle
\section{Introduction}

Photometric redshifts have been regarded as the most promising
tool in studying the formation and evolution of galaxies and the
large scale structure of the universe, especially when the spectra
of faint objects are difficult to obtain. The photometric redshift
technique translates observables such as flux and apparent color
to the corresponding intrinsic properties of absolute luminosity
and rest-frame color. The idea behind the photometric redshift
technique is to measure the redshifts of galaxies and AGN based on
available multi-wavelength photometry. Photometric redshift
techniques can be traced back to Baum (1962) who used nine
medium-wide filters to detect the 4000{\AA} in galaxies. For
example, the predicted redshft of the C10925 galaxies by this
technique is $z = $0.19, which agrees closely with the known
spectroscopic value of $z = $0.192. Subsequent implements have
been made by Koo (1985) using four broad-band photographic
filters, Loh and Spillar (1986) using CCDs along with 6
medium-band filters, and Xia et al. (2002) using CCD photometries
of BATC 15 medium-band filters. In the last two decades, some
well-defined statistical techniques have become increasingly
popular in predicting photometric redshifts.

There are two kinds of different photometric redshift approaches in
the astronomical literature: the template fitting whose templates
are derived from synthetic (e.g. Bruzual, Charlot 1993) or empirical
template spectra (e.g. Coleman, Wu, Weedman 1980), and the empirical
training set method which constructs a direct empirical correlation
between color and redshifts. For template fitting one, according to
the known redshift and galaxy type, some templates are constructed
in advance. By minimizing the standard $\chi^{2}$ to fit the
observed photometric data with a set of spectral templates, this
method can be applied beyond redshift limit. Although it is easy to
implement, the accuracy of this approach strongly depends on the
templates. The essence of training set approach is to derive a
function between redshift and photometric data by using a large and
representative training set of galaxies for which both photometry
and redshift are known, and then use this function to estimate the
redshifts of objects with unknown redshifts. In the last few years,
a large number of training set methods have been developed and used
(Way, Srivastava 2006). For example, linear or non-linear fitting
(Brunner et al. 1997; Wang, Bahcall, Turner 1998; Budavari et al.
2005); support vector machines (SVMs, Wadadekar 2005); artificial
neural network (ANNs, Firth, Lahav, Somerville 2003, Ball et al.
2004, Collister, Lahav 2004, Vanzella et al. 2004, Li et al. 2006);
instance-based learning (Csabai et al. 2003; Ball et al. 2007).

The main advantage of SVMs over ANNs is that requires less effort in
training, and no danger of overfitting. SVMs simplify the decision
of the optimal networks by replacing the choice of architecture
problem with one of choice of kernel (Wadadekar 2005). The strength
of instance-based learning methods is that they needn't training,
but implement their predictions directly on (training) data that has
been stored in the memory. In general, they store all the training
data in the memory during the learning phase, and defer all the
essential computation until the prediction phase. Kernel regression
(KR) belongs to the instance-based learning family. Based on the
merits of SVMs and KR, we adopt these two methods to predict
photometric redshifts of galaxies.

In this paper we explore two approaches: support vector machines
(SVMs) and kernel regression (KR) to estimate redshifts of galaxies
with photometric data from SDSS and 2MASS databases. The structure
of this paper is as follows: Section 2 illustrates the data used in
the study. Section 3 describes the principles of SVMs and KR.
Section 4 gives the results and discussion. Finally the conclusions
are summarized in Section 5.

\section{Data}
The data used in this paper is from Sloan Digital Sky Survey (SDSS)
and Two Micron All Sky Survey (2MASS) catalogs. The process of data
preprocessing has been done by VO\_DAS, which is a data accessing
system of Virtual Observatory of China. The general information of
SDSS and 2MASS is shown as follows.

The Sloan Digital Sky Survey (SDSS, York et al. 2000) is an
astronomical survey project, which covers more than a quarter of
the sky, to construct the first comprehensive digital map of the
universe in 3D, using a dedicated 2.5-meter telescope located on
Apache Point, New Mexico. In its first phase of operations, it has
imaged 8,000 square degrees in five bandpasses ($u, g, r, i, z$)
and measured more than 675,000 galaxies, 90,000 quasars and
185,000 stars. In its second stage, SDSS will carry out three new
surveys in different research areas, such as the nature of the
universe, the origin of galaxies and quasars and the formation and
evolution of the Milky Way.

The Two Micron All Sky Survey (2MASS, Cutri et~al. 2003) uses two
highly-automated 1.3-m telescopes, one is at Mt. Hopkins, Arizona
and the other locates at CTIO, Chile. Each telescope is equipped
with a three-channel camera, each channel consisting of a 256x256
array of HgCdTe detectors, capable of observing the sky
simultaneously at $J$ (1.25\,$\mu$m), $H$ (1.65\,$\mu$m), and $Ks$
(2.17\,$\mu$m), to a 3$\sigma$ limiting sensitivity of 17.1, 16.4
and 15.3\,mag in the three bands. Jarrett et al. (2000) has more
detailed information on the extended source catalog.

We select all galaxies with known spectra redshifts from SDSS Data
Release 5, and then cross-match the data with 2MASS extended source
catalog within a search radius of 3 times the SDSS positional
errors. After cross-matching, we generate about 150,000 galaxies.
From these galaxies, we selected the objects satisfying the
following criteria: 1) the spectroscopic redshift confidence must be
equal to or greater than 0.95; 2) redshift warning flag is 0; 3) r
$< 17.5$. With these qualifications, a sample of 62,083 galaxies is
obtained. Table 1 describes the broadband filters and their
wavelength range from SDSS and 2MASS catalogs.

\begin{table*}[ht]
\begin{center}
\caption{ Survey filters and characteristics}
\begin{tabular}{llll}
\hline \hline
Bandpass & Survey & $\lambda_{eff}({\AA})$ & $\Delta\lambda({\AA})$\\
\hline
$u$ & SDSS & 3551 & 600\\
$g$ & SDSS & 4686 & 1400\\
$r$ & SDSS & 6165 & 1400\\
$i$ & SDSS & 7481 & 1500\\
$z$ & SDSS & 8931 & 1200\\
$J$ & 2MASS & 12500 & 1620\\
$H$ & 2MASS & 16500 & 2510\\
$Ks$ & 2MASS & 21700 & 2620\\
\hline
\end{tabular}
\bigskip
\end{center}
\end{table*}

\section{Model Selection}

\subsection{Support Vector Machines}

The foundation of Support Vector Mahines (SVMs) has been developed
by Vapnik (1995). SVMs were developed to solve the classification
problem, but recently they have been extended to the domain of
regression problems. The regression problem of SVMs is achieved by
using an alternative loss function, which must be modified to
include a distance measure. The SVM task usually involves with
training and testing data which consist of some data instances.
Each instance in the training set contains one ``target value" and
several ``attributes". The goal of SVMs is to produce a model
which predicts target value of data instances in the testing set
which are given only the attributes.

Given a training set of training pairs $(x_{1},y_{1})$,...
,$(x_{l},y_{l})$, $x_{i}\in R^{n}$, $y \in R$, with a linear
function,
\begin{equation}
f(x) = <\omega,x>+b,
\end{equation}
The optimal regression function is given by the minimum of the
functional,
\begin{equation}
\phi(\omega,\zeta)=\frac{1}{2}\omega.\omega+C\sum_{i}(\zeta_{i}^{-}+\zeta_{i}^{+}),
\end{equation}
Using a quadratic loss function,
\begin{equation}
L_{quad}(f(x)-y)=(f(x)-y)^{2},
\end{equation}
the solution is given by,

\begin{displaymath}
\max_{\alpha,\alpha^{*}}W(\alpha,\alpha^{*})=\max_{\alpha,\alpha^{*}}
-\frac{1}{2}\sum_{i=1}^{l}\sum_{j=1}^{l}(\alpha_{i}-\alpha_{i}^{*})(\alpha_{j}-\alpha_{j}^{*})<x_{i},x_{j}>
\end{displaymath}
\begin{equation}
+\sum_{i=1}^{l}(\alpha_{i}-\alpha_{i}^{*})y_{i}
-\frac{1}{2C}\sum_{i=1}^{l}(\alpha_{i}^{2}+(\alpha_{i}^{2})^{2}),
\end{equation}

the resultant optimization problems is,
\begin{equation}
\min_{\beta}\frac{1}{2}\sum_{i=1}^{l}\sum_{j=1}^{l}\beta_{i}\beta_{j}<x_{i},x_{i}>-\sum_{j=1}^{l}\beta_{i}
y_{i}+\frac{1}{2C}\sum_{i=1}^{l}\beta_{i}^{2}
\end{equation}
 with constraints,
 \begin{equation}
 \sum_{i=1}^{l}\beta_{i}=0.
 \end{equation}
To generalize to non-linear regression, we replace the dot product
with a kernel function. More information can be found in Steve's
tutorial (1998). In our work, we adopt the Gaussian kernel
function.

SVMs have been widely used in the area of machine learning due to
its excellent generalization performance, such as handwritten digit
recognition and face detection. In astronomy, SVMs have been applied
to identifying red variables (Williams et al. 2004), clustering
astronomical objects (Zhang, Zhao 2004), and classifying AGN from
stars and normal galaxies (Zhang, Cui, Zhao 2002).

Several software implementations of the SVM algorithm are
accessible on the web. Due to their robustness, the ability of
handling large amounts of data, and the regression time, we use
SVM\_Light for our study. SVM\_Light is fast optimized SVM
algorithm, which is implemented in C language. It can deal with
many thousands of support vectors, handle hundreds/thousands of
training examples, and provide several standard kernel functions.
The details about SVM\_Light can be found at
http://www.cs.cornell.edu/People/tj/svm\_light/.

\subsection{Kernel Regression}

Kernel regression (KR) belongs to the family of instance-based
learning algorithms (Watson 1964; Nadaraya 1964), which simply
store some or all of the training examples and do not perform any
kind of generalization of the given samples and ``delay learning"
till prediction time. Given a query point $x_q$, a prediction is
obtained using the training samples that are ``most similar" to
$x_q$. Similarity is measured by means of a distance metric
defined in the hyper-space of $V$ predictor variables. Kernel
regressors obtain the prediction for a query point $x_q$, by a
weighted average of the $y$ values of its neighbors. The weight of
each neighbor is calculated by a function of its distance to $x_q$
(called the kernel function). These kernel functions give more
weight to neighbors that are nearer to $x_q$. The notion of
neighborhood (or bandwidth) is defined in terms of distance from
$x_q$. The prediction for query point $x_q$ is obtained by
\begin{equation}
y_q={\frac{\sum\limits_{i=1}^{N}K({\frac{D(x_i,x_q)}{h})}\times
y_i}{{\sum\limits_{i=1}^{N}K({\frac{D(x_i,x_q)}{h}})}}}
\end{equation}
where $D(.)$ is the distance function between two instances; $K(.)$
is a kernel function; $h$ is a bandwidth value; $(x_i, y_i)$ are
training samples. In this paper, we use Euclidian distance and
Gaussian kernel function. $x_i$ is the feature for each training
sample, $y_i$ is the spectroscopic redshift for each training set
sample, $y_q$ is the redshift of each query sample.

One important design decision when using kernel regression is the
choice of the bandwidth $h$. The larger $h$ results in the flatter
weight function curve, which indicates that many points of training
set contribute quite evenly to the regression. As the $h$ tends to
infinity, the predictions approach the global average of all points
in the database. If the $h$ is very small, only closely neighboring
data points make a significant contribution. If the data are
relatively noisy, we expect to obtain smaller prediction errors with
a relatively larger $h$. If the data are noise free, then a small
$h$ will avoid smearing away fine details in the function. There
exists mature algorithms for choosing the bandwidth for kernel
regression that minimize a statistical measure of the difference
between the true underlying distribution and the estimated
distribution. Usually bandwidth selection in regression is done by
cross-validation (CV).

In this work, we chose the bandwidth using cross-validated method.
Cross-validation is the statistical method of dividing a sample of
data into subsets such that the analysis is initially performed on a
single subset, while the other subset(s) are retained for subsequent
use in confirming and validating the initial analysis. $M$-fold
cross-validation is one important cross-validation method. The data
are divided into $m$ subsets of (approximately) equal size. Each
time, one of the $m$ subsets is used as the test set and the other
$m-1$ subsets are put together to form a training set for a given
bandwidth. Then the average error across all $m$ trials is computed
(Zhang, Zhao 2007). Here we adopt 10-fold cross-validation for the
bandwidth choice dividing the samples into 10 subsets, then 9
subsets of 10 subsets are taken as training set and the rest subset
as testing set for ten times. The optical bandwidth is indicated by
the bandwidth with the minimum of average errors. In Table 2, we
apply KR with 7-color ($u-g, g-r, r-i, i-z, z-J, J-H, H-Ks$) and
spectra redshifts as an input pattern, taking it as an example to
illustrate the relationship between bandwidth ($h$) and
cross-validated value (CV). It is obvious that the optimal bandwidth
$h$ is 0.045 when cross-validated value arrives at the minimum 4.33.

\begin{table*}[ht]
\begin{center}
\caption{ The relationship between bandwidth ($h$) and
cross-validated value (CV).}
\begin{tabular}{rllllllllll}
\hline \hline
$h$   & 0.015 & 0.02 & 0.025 & 0.03 & 0.035 & 0.04 & $\textbf{0.045}$ & 0.05 & 0.055 & 0.06 \\
\hline
CV($\times10^{-5}$) & 4.77 & 5.03 & 4.86 & 4.64 & 4.45 & 4.35 & $\textbf{4.33}$ & 4.35 & 4.41 & 4.77\\

\hline
\end{tabular}
\bigskip
\end{center}

\end{table*}

\section{Result and Discussion}
One advantage of the empirical training set approach to
photometric redshift estimation is that additional parameters can
be easily incorporated. More parameters (e.g. $petro50\_r$,
$petro90\_r$, $fracDeV\_r$, etc.) may be taken as inputs. In order
to study which parameters influence the accuracy of predicting
photometric redshifts, we probe different input patterns to
estimate photometric redshifts. We randomly divide the sample into
two parts: 41,388 for training and 20,695 for test, and apply them
to train and test kernel regression (KR) and support vector
machines (SVMs). The rms deviations of predicting photometric
redshifts with KR and SVMs for different situations are listed in
Table 3.

When using SVMs to estimate photometric redshifts, the performances
of colors are the best of all cases. The result based on 4-color
input pattern ($u-g,g-r,r-i,i-z$) has the same best accuracy
($\sigma_{\rm rms}$=0.0273) as that based on 7-color input pattern
($u-g,g-r,r-i,i-z,z-J,J-H,H-Ks$). The accuracy with seven colors and
$r$ magnitude is better than that with four colors and $r$
magnitude, and than that with seven magnitudes. The performance
taking five magnitudes as input is not as well as that of seven
magnitudes. Since the accuracy with four colors is best, we consider
more parameters besides four colors to probe whether the performance
improves. As shown by Table 3, the accuracy adding $fracDev\_r$ or
$petro50\_r$ and $petro90\_r$ decreases. Obviously the more
parameters considered, the performance is not always better,
sometimes even worse. The results have shown that there is no
improvement in the use of more parametric data from SDSS and 2MASS
catalogs. Hence we would not recommend its use because it decreases
the sample size markedly and does not decrease the rms errors in the
photometric redshift prediction.

\begin{table*}[ht]
\begin{center}
\caption{The dispersions of photometric redshift prediction using
KR and SVMs.}
\begin{tabular}{lll}
\hline \hline
  method &                  KR &  SVMs  \\
\hline
Input Parameters&$\sigma_{\rm rms}$(optimal bandwidth)&$\sigma_{\rm rms}$\\

\hline
$u,g,r,i,z$                            &0.0208 ($h=0.025$) &0.0291\\
$u,g,r,i,z,J,H,Ks$                      &0.0254 ($h=0.015$) &0.0278\\
$u-g,g-r,r-i,i-z$                      &0.0193 ($h=0.020$) &0.0273\\
$u-g,g-r,r-i,i-z,r$                    &0.0196 ($h=0.025$) &0.0284\\
$u-g,g-r,r-i,i-z,z-J,J-H,H-Ks$          &0.0210 ($h=0.045$) &0.0273\\
$u-g,g-r,r-i,i-z,z-J,J-H,H-Ks,r$        &0.0235 ($h=0.055$) &0.0275\\
$u-g,g-r,r-i,i-z,fracDev\_r$           &0.0192 ($h=0.020$) &0.0306\\
$u-g,g-r,r-i,i-z,petro50\_r,petro90\_r$&0.0218 ($h=0.040$) &0.0330\\
\hline
\end{tabular}
\bigskip
\end{center}
NOTE.----$petro50\_r$ is Petrosian 50\% radius in $r$ band,
$petro90\_r$ is Petrosian 90\% radius in $r$ band, $fracDeV\_r$ is
$fracDeV$ in $r$ band.
\end{table*}

For KR, the best input patterns includes four colors and
$fracDev\_r$ ($u-g$, $g-r$, $r-i$, $i-z$, $fracDev\_r$) when the rms
error amounts to 0.0192. The better input patterns are four colors
or four colors and r magnitude when rms error are 0.0193 or 0.0196,
respectively. Then the good input pattern includes five magnitudes
when the rms scatter is 0.0208. The result with only seven colors is
better than that with seven colors and r magnitude but worse than
that with five magnitudes. For four colors as inputs, the
performance of kernel regression decreases when adding $petro50\_r$
and $petro90\_r$, except for $fracDev\_r$. As a result, when
applying kernel regression to predicting photometric redshifts, we
find the parameters except magnitudes and color indexes, such as
$petro50\_r$ and $petro90\_r$, are void, however $fracDev\_r$ is
important and effective possibly because $fracDev\_r$ is closely
related to galaxy type. When implementing KR, the enlargement of
bandwidth may cause less loss of estimation. In our experiments, the
fraction of loss is less than 1\%. Table 3 also indicates that the
optimal bandwidth is different for different input patterns.

To clearly show the performances of KR and SVMs, we take the best
and worst results for both approaches, and plot the known
spectroscopic redshifts against the calculated photometric redshifts
from the test data, shown in Figures 1-2. Figure 1 depicts the
results for KR, while Figure 2 gives the results for SVMs. In
Figures 1 and 2, the left panel shows the best input pattern, and
the right panel plots the worst input pattern. It is clear that, for
SVMs, the estimation of photometric redshifts is high for
low-redshift galaxies. However, the performance of KR is very
satisfactory.

\begin{figure}[h!!!]
\includegraphics[scale=0.5]{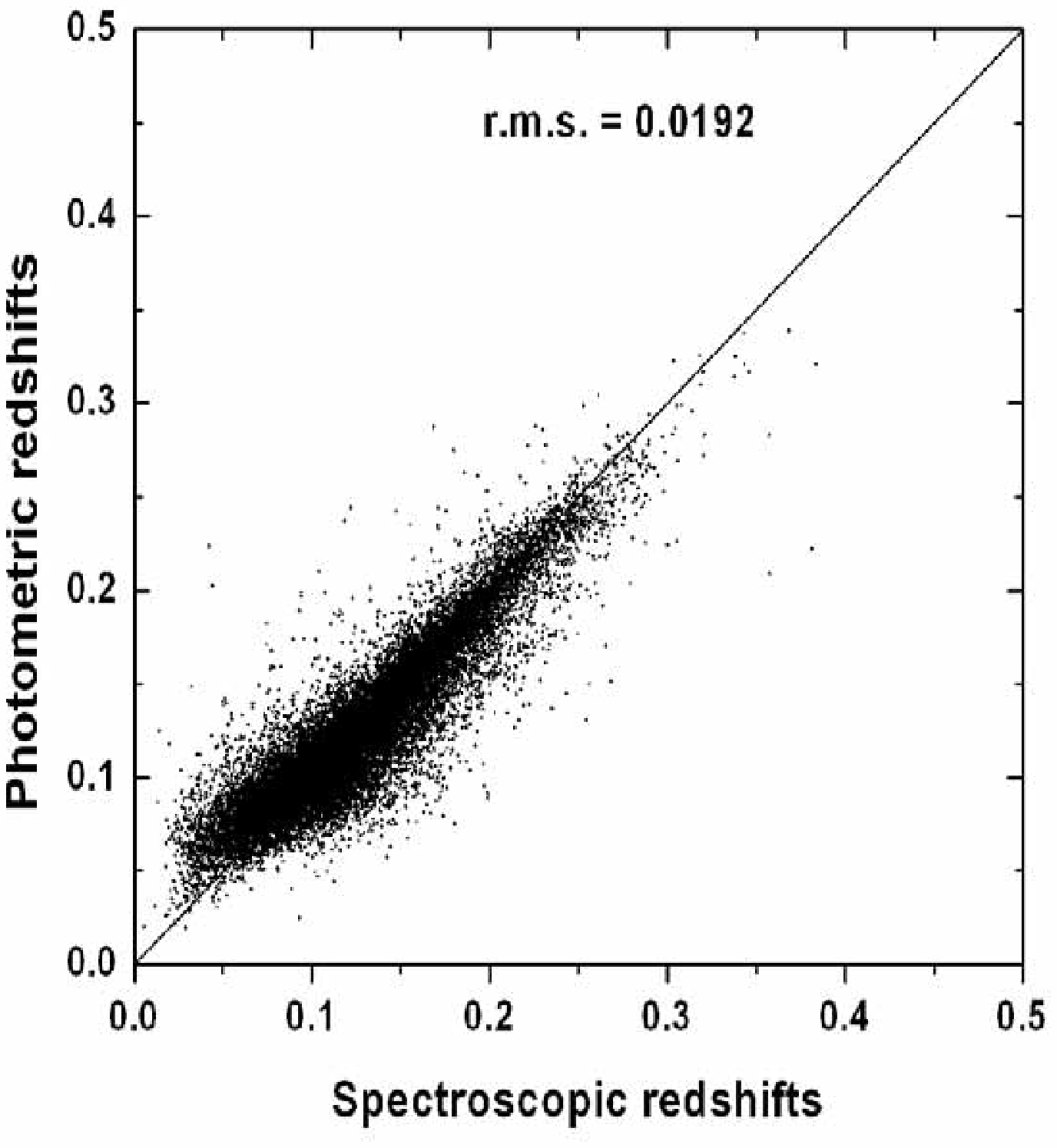}
\includegraphics[scale=0.5]{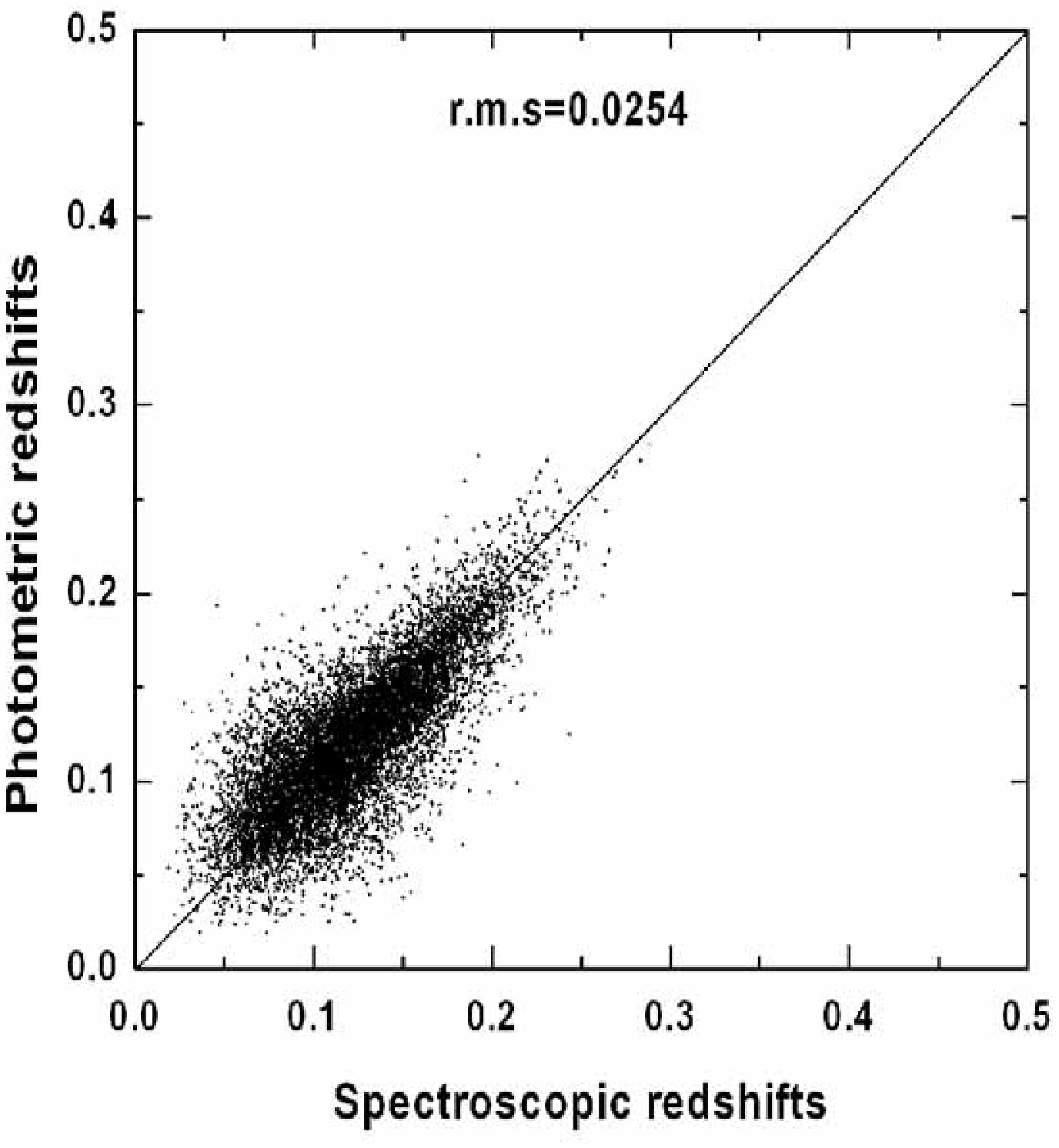}
\caption[]{Spectroscopic redshift versus calculated photometric
redshift comparisons using 20,695 test galaxies from the SDSS DR5
and 2MASS databases with kernel regression. Left figure shows that
the best input pattern with $\sigma_{rms}$=0.0192 is
$u-g,g-r,r-i,i-z,fracDev\_r$. Right figure indicates that the worst
input pattern with $\sigma_{rms}$=0.0254 is $u,g,r,i,z,J,H,Ks$.}
 \label{fig1}
\end{figure}

\begin{figure}[h!!!]
\includegraphics[scale=0.5]{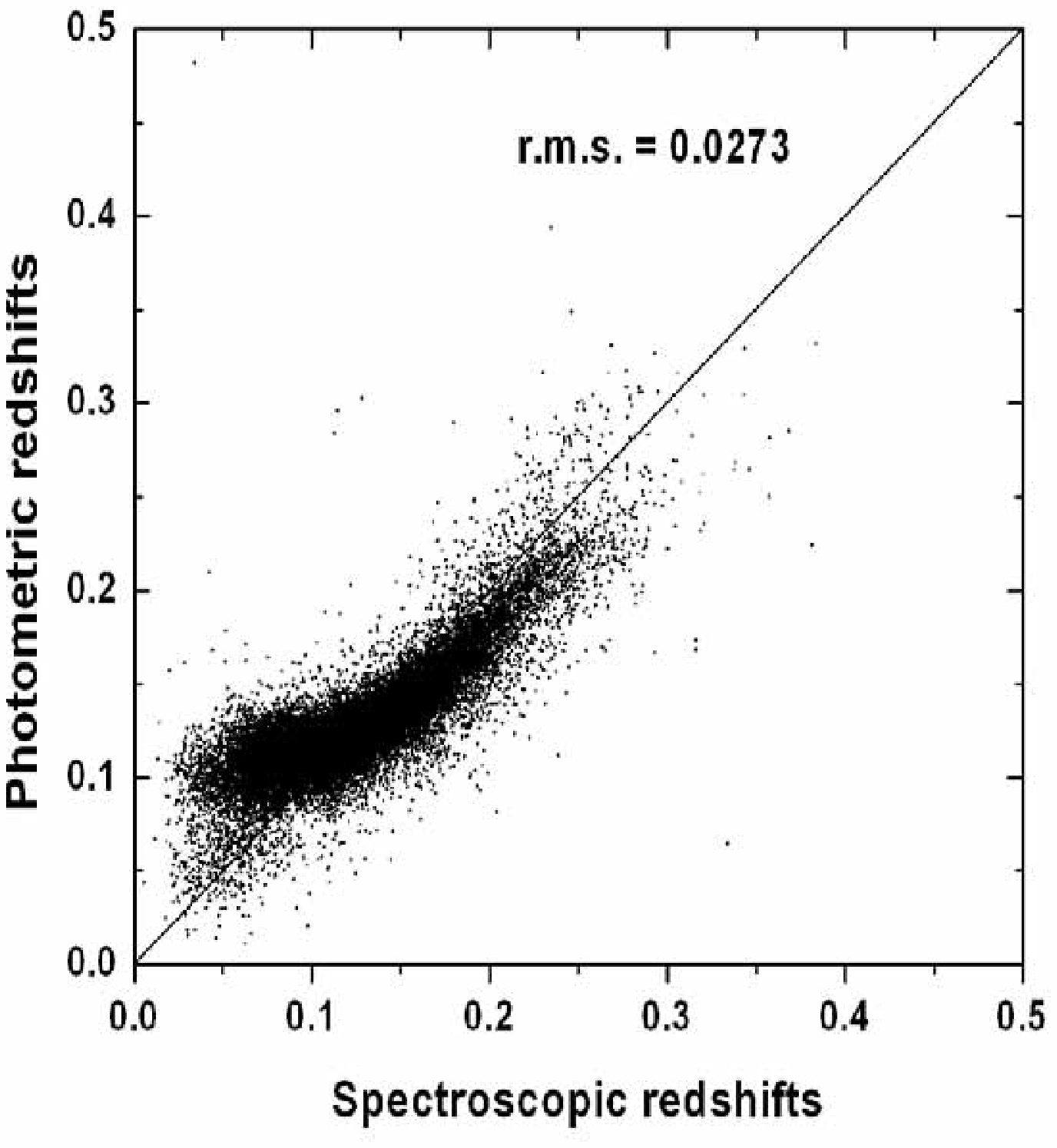}
\includegraphics[scale=0.5]{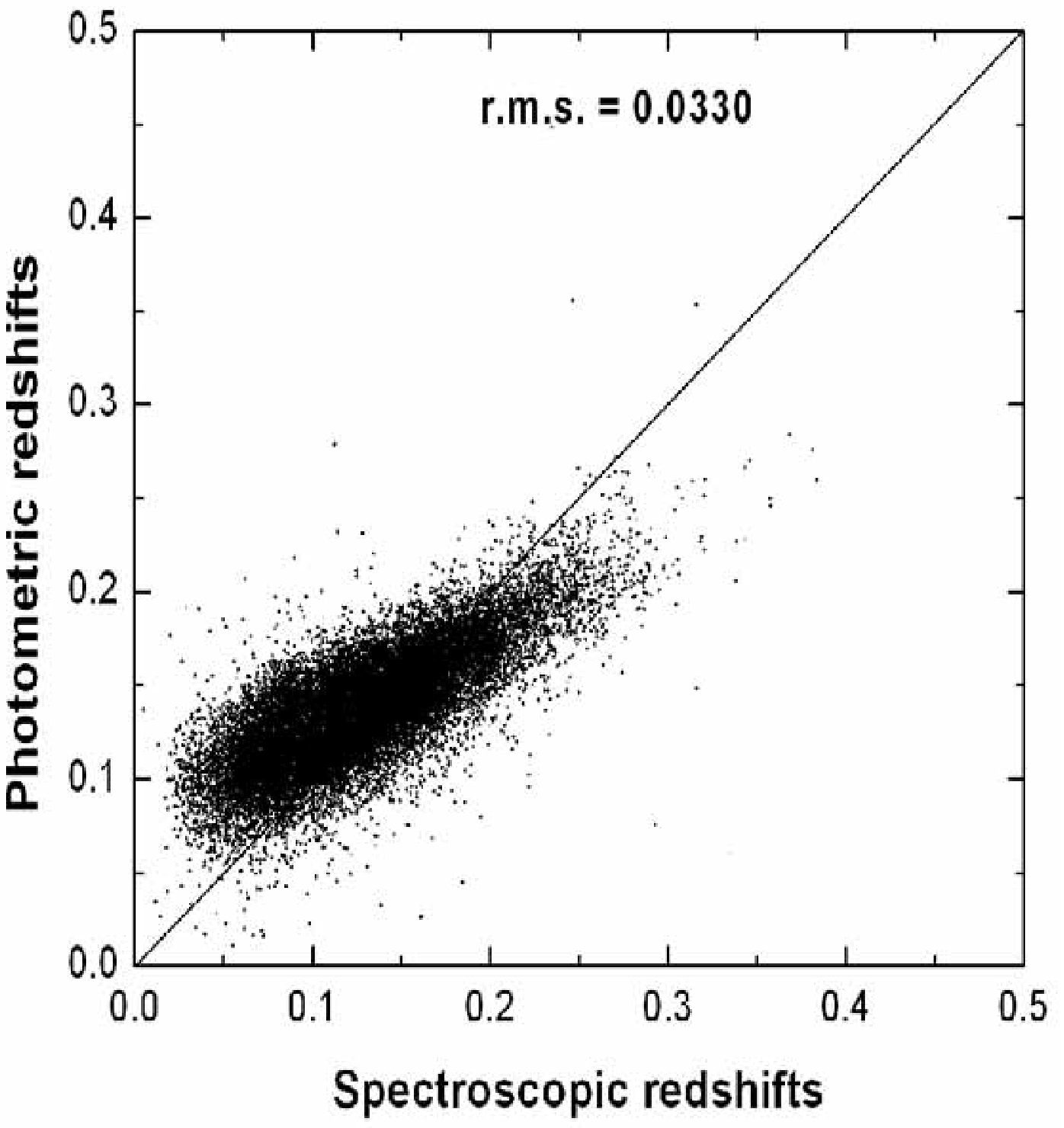}
\caption[]{Spectroscopic redshift versus calculated photometric
redshift comparisons using 20,695 test galaxies from the SDSS DR5
and 2MASS databases with SVMs. Left figure shows that the best input
pattern with $\sigma_{rms}$=0.0273 is $u-g,g-r,r-i,i-z$. Right
figure indicates that the worst input pattern with
$\sigma_{rms}$=0.0330 is $u-g,g-r,r-i,i-z,petro50\_r,petro90\_r$. }
 \label{fig1}
\end{figure}

So far there has been much work on approaches to photometric
redshift estimation. To compare the performance of various methods,
we list the rms scatters of photometric redshifts from different
work in Table 4. Because the accuracy strongly depends on the data
used, we only give a coarse comparison. As shown in Table 4, kernel
regression is comparable to artificial neural networks (ANNs),
better than SVMs (Wadadekar 2005), Kd-tree (Csabai et al. 2003) and
polynomial (Connolly et al. 1995), and superior to CWW and
Bruzual-Charlot (Csabai et al. 2003). Nevertheless, each method has
its strongness and weakness. Kernel regression belongs to the
instance-based learning family. It is a kind of memory-based method
and learns until prediction. Therefore kernel regression consumes
much large memory of a computer in despite of high accuracy. If
using ANNs, one should be familiar with the network architecture and
make a decision about how many input nodes or hidden lays they have.
The more complex networks it has, the more accurate result it earns.
However, SVMs may use different kernel functions instead of
different ANN networks. As long as you choose the appropriate kernel
function and parameters, the rms scatter will decrease
significantly. Moreover the classical problems such as multi-local
minima, curse of dimensionality and overfitting in ANNs, seldom
occur in SVMs. Nevertheless, SVMs need prior knowledge to adjust
parameters. Degeneration between parameters makes the regulating
process more complicated. Even though linear or non-linear
polynomial regression is easy to implement and communicate with
astronomers, the systematic deviation is large (Brunner et al. 1997;
Wang et al. 1998; Budav$\acute{a}$ri et al. 2005; Hsieh et al. 2005;
Connolly et al. 1995). Csabai et al. (2000) have represented a
hybrid method, which is a combination of template-based and
empirical training set. The hybrid one can reconstruct the continuum
spectra of galaxies directly from a set of multicolor photometric
observations and spectroscopic redshifts. Even though the dispersion
of photometric redshifts using this combination technique was
significantly improved, it is still worse than empirical ones.

\begin{table*}[ht]
\begin{center}
\caption{Photometric redshift accuracies for various approaches}
\begin{tabular}{llllll}
\hline \hline
Method Name & $\sigma_{\rm rms}$ & Data set & Input parameters \\
\hline
CWW$^{1}$         &  0.0666    & SDSS-EDR & $ugriz$ \\
Bruzual-Charlot$^{1}$    &  0.0552    & SDSS-EDR & $ugriz$  \\
Interpolated$^{1}$&0.0451& SDSS-EDR & $ugriz$\\
Polynomial$^{1}$    &  0.0318    & SDSS-EDR & $ugriz$ \\
Kd-tree$^{1}$      &  0.0254 & SDSS-EDR & $ugriz$  \\
ClassX$^{2}$  &  0.0340    & SDSS-DR2 & $ugriz$  \\
SVMs$^{3}$    &  0.0270 & SDSS-DR2 & $ugriz$   \\
ANNs$^{4}$       &  0.0229   & SDSS-DR1 & $ugriz$ \\
Polynomial$^{5}$     & 0.0250 & SDSS-DR1,GALEX & $ugriz+nuv$ \\
Kernel Regression & 0.0208 & SDSS-DR5,2MASS & $ugriz$  \\
   & 0.0193 & SDSS-DR5,2MASS & $color^{*}$  \\
SVMs & 0.0273 & SDSS-DR5,2MASS & $color^{*}$ \\
\hline
\end{tabular}
\bigskip
\end{center}
NOTE.---- SDSS-EDR = Early Data Release (Stoughton et al.
2002),\\SDSS-DR1 = Data Release 1 (Abazajian et al. 2003),\\
SDSS-DR2 = Data Release 2 (Abazajian et al. 2004),\\ SDSS-DR5 = Data
Release 5 (Adelman-McCarthy et al. 2007).\\$color^{*}$  is the color
indexes, i.e. $u-g$, $g-r$, $r-i$, $i-z$. \\ (1) Csabai et al. 2003;
(2) Suchkov, Hanisch, Margonet 2005;\\(3) Wadadekar 2005; (4)
Collister, Lahav 2004; (5) Budav$\acute{a}$ri et al. 2005.
\end{table*}

\section{Conclusions}
We utilize two novel methods, which are Support Vector Machines
(SVMs) and Kernel Regression (KR), to estimate photometric
redshifts using the cross-matched data from SDSS DR5 and 2MASS. We
compare the performances of estimating photometric redshifts with
SVMs and KR for different input patterns. Our experiments show
that only when the appropriate parameters are chosen, the accuracy
of SVMs or KR can improve. Adding additional bandpasses from the
infrared (2MASS) contribute little information due to the small
size of dataset. In addition, there is no improvement in the use
of the parameters ($petro50\_r,petro90\_r,fracDev\_r$) related to
angular size and morphology.

The accuracy of photometric redshift produced by SVMs is slightly
less than that of ANNs, as good as linear or quadratic regression,
and clearly much better than template fitting one. In appropriate
situations, SVMs will be a highly competitive tool for determining
photometric redshifts in terms of speed and applications. However,
it does depend on the existence of a large and representative
training sample. As a kind of empirical photometric redshift
estimations, SVMs are impossible in extrapolating to the region that
is not well sampled by training set. Moreover a potential solution
to the problem of increasing the photometric redshift accuracy is to
choose the more appropriate kernel function, and to consider the
feature selection/extraction methods in the process of parameter
selection.

The dispersion of photometric redshift estimation by kernel
regression is fairly favorable. Compared to other training-set
methods, kernel regression does not need any effort on training.
In addition, kernel regression ameliorates a major problem of
empirical training-set methods. Even though a few high-redshift
galaxies exists in the sample, kernel regression can appropriately
adjust bandwidth to obtain much more accurate redshifts.
Therefore, kernel regression can extrapolate to regions where the
input parameters are not well represented by the training data.
With large and deep photometric surveys carried out, it seems that
kernel regression will show its superiority. In the future work we
will explore adaptive bandwidth or other kinds of distance metric
for kernel regression on the regression problems.

\begin{acknowledgements}
The authors acknowledge the referee whose insightful and detailed
comments improved the presentation of this paper. This research has
made use of data products from the Two Micron All Sky Survey, which
is a joint project of the University of Massachusetts and the
Infrared Processing and Analysis Center/California Institute of
Technology, funded by the National Aeronautics and Space
Administration and the National Science Foundation.

Funding for the SDSS and SDSS-II has been provided by the Alfred
P. Sloan Foundation, the Participating Institutions, the National
Science Foundation, the U.S. Department of Energy, the National
Aeronautics and Space Administration, the Japanese Monbukagakusho,
the Max Planck Society, and the Higher Education Funding Council
for England.

The SDSS is managed by the Astrophysical Research Consortium for
the Participating Institutions. The Participating Institutions are
the American Museum of Natural History, Astrophysical Institute
Potsdam, University of Basel, Cambridge University, Case Western
Reserve University, University of Chicago, Drexel University,
Fermilab, the Institute for Advanced Study, the Japan
Participation Group, Johns Hopkins University, the Joint Institute
for Nuclear Astrophysics, the Kavli Institute for Particle
Astrophysics and Cosmology, the Korean Scientist Group, the
Chinese Academy of Sciences (LAMOST), Los Alamos National
Laboratory, the Max-Planck-Institute for Astronomy (MPIA), the
Max-Planck-Institute for Astrophysics (MPA), New Mexico State
University, Ohio State University, University of Pittsburgh,
University of Portsmouth, Princeton University, the United States
Naval Observatory, and the University of Washington.

This paper is funded by National Natural Science Foundation of
China under grant No.90412016, 60603057 and 10778623.
\end{acknowledgements}

\label{lastpage}

\end{document}